\documentclass[aps,twocolumn,groupedaddress,prb,final]{revtex4}
\usepackage{epsfig}
\usepackage{amsmath}

\preprint{Submitted to \textsl{Computer Physics Communications}}
\newcommand{\psfile}[2]{\epsfxsize=#2\linewidth\centerline{\epsfbox{#1}}}
 \newcommand{\Rv}{{\bf R}}

 \newcommand{\Fv}{{\bf F}}
 \newcommand{\rv}{{\bf r}}

\begin{document}

\title{HARES: an efficient method for first-principles electronic structure 
       calculations of complex systems}
\author{
U.~V.~Waghmare$^{(1)}$, Hanchul Kim$^{(2)}$, I.~J.~Park$^{(2)}$,
Normand Modine$^{(3)}$, P.~Maragakis$^{(2)}$, Efthimios Kaxiras$^{(1,2)}$} 

\affiliation{$^{(1)}$ Department of Physics, Harvard University, Cambridge, MA 02138}
\affiliation{$^{(2)}$ Division of Engineering and Applied Sciences}
\affiliation{Harvard University, Cambridge, MA 02138}
\affiliation{$^{(3)}$ Sandia National Laboratory, Albuquerque, NM 87185}


\begin{abstract}
We discuss our new implementation of the 
Real-space Electronic Structure method for 
studying the atomic and electronic structure of infinite periodic 
as well as finite systems, based on density functional theory. 
This improved version which we call HARES (for
High-performance-fortran Adaptive grid Real-space Electronic
Structure) 
aims at making the method widely applicable and 
efficient, using high performance Fortran on parallel architectures.
The scaling of various parts of a HARES calculation 
is analyzed and compared to that of plane-wave based methods. 
The new developments that lead to enhanced performance, and their
parallel implementation, are presented in detail.  
We illustrate the application of HARES to the study of elemental 
crystalline solids, molecules and 
complex crystalline materials, such as blue bronze and zeolites.
\end{abstract}

\maketitle

\section{Introduction}
\label{intro} 

Technological advances are pushing the size of device components and the demands 
on their performance to ever smaller sizes and higher standards.
These trends, which are only expected to accelerate in the future,
make it imperative that the structure and behavior of systems at the 
atomistic level are thoroughly understood from a fundamental perspective.  
While experimental methods are steadily improving in their ability
to probe atomistic processes of materials, 
computational approaches provide a complementary technique for
systematic and well-controlled studies. Tuning and optimization of 
the properties of materials, taking into account their chemical composition, 
require computational methods that are 
of general applicability, unbiased and accurate.

First-principles methods based on density functional theory (DFT) \cite{dft}
have proven to be an accurate and reliable tool in understanding and predicting
a wide range of physical properties of finite 
(such as molecules and clusters) and extended structures (such as 
bulk crystalline solids, their defects and surfaces). 
Such
methods must obtain the quantum mechanical ground state of the interacting 
electrons and ions, which makes them 
computationally very intensive. The computational cost scales as a rather 
high power (typically 3) of  
the number of atoms or electrons in the system, which limits 
the sizes that
can be investigated in a reasonable time.  Improvement of the
efficiency of first-principles calculations is therefore an important goal. 
This goal can 
be reached either by improving the algorithms to obtain
better scaling with system size, or by exploiting modern computational 
resources and in particular parallel architectures.

The scaling of the calculation of the quantum mechanical ground state of 
interacting electrons and ions can be improved by exploiting what W. Kohn
has called ``the short-sightedness'' of quantum mechanics: due to 
screening, interactions are essentially  
{\it short} ranged.  A natural way to express this property is through 
the density matrix of the system.  Thus, implementations of 
linear scaling (referred to as $O(N)$ methods, with $N$ 
a number representing the system size, like the number of electrons) 
typically 
involve density-matrix expressions with
localized orbitals~\cite{linear}. In terms of the
computational time, $O(N)$ methods prove advantageous for 
insulating systems with $N > 10^{3}$ or metallic systems with $N > 10^{4}$ 
electrons
(the number of atoms in the system is typically an order of 
magnitude smaller than the number of electrons).  
This difference in efficiency of the $O(N)$ methods between insulators
and metals is related to the long-range behavior of the density matrix 
which has a different fall off, i.e. exponential in the former
vs. power law in the latter.
These methods are well suited for the calculation of the total energy 
of the system, which provides useful information about its optimal 
structure, dynamics, and response to mechanical loading. 
In addition to the total energy, it is often important to study 
the electronic structure of the system.  This is necessary for 
understanding electronic, optical and magnetic properties,
and is relevant for the study of both insulators (semiconductors) and metals.  
A DFT electronic structure calculation requires the calculation 
of the eigenvalue
spectrum of the single-particle Hamiltonian, 
a problem which is not easily amenable to improvements in scaling since 
diagonalization of the Hamiltonian typically scales as $N^3$.
Localization of the electronic orbitals can be detrimental to the accuracy
of electronic properties 
though it helps improve the efficiency of $O(N)$
methods.

In DFT calculations, the single-particle Hamiltonian matrix itself 
depends on the eigenfunctions, so the complete solution 
must be obtained by iterating the solution to self-consistency. 
The size of the Hamiltonian matrix depends on the basis set 
used to represent the 
electronic wavefunctions and the electronic charge density. 
Ideally, one would like to work with 
a sparse Hamiltonian, which can
be solved efficiently using iterative algorithms; 
the use such algorithms 
reduces both computer memory and time requirements. 
A natural way to generate a sparse Hamiltonian is to use a real 
space grid for the representation of the 
electronic eigenfunctions and charge density: 
each term in the Hamiltonian, evaluated at some point in space, 
acts only on the wavefunctions at the same point in space,
except for the Laplacian in the kinetic energy operator which 
involves several points simultaneously.  The number of points  
included in the evaluation of the Laplacian determines the few 
off-diagonal non-zero
matrix elements in each row (or column) of the Hamiltonian matrix.  
The calculation can be made even more efficient by using an adaptive 
grid in real space for  
representation of the eigenfunctions, with points distributed 
according to the electronegativity of ions.       
The adaptive grid can be mapped onto a regular grid 
in curvilinear space through the proper definition 
of a metric.              
The regular grid in curvilinear space makes it possible to 
exploit fully the capabilities of 
modern computational platforms based on parallel processing. 
Thus, this formulation of the problem satisfies all requirements 
for very efficient electronic 
structure calculations: (a) {\em Sparsity} for fast iterative 
diagonalization of the Hamiltonian; (b) {\em Adaptability} for 
efficient distribution of grid points as demanded by the 
physical system; and (c) {\em Efficient parallelization} of the 
computation due to the natural distribution of the regular 
curvilinear space grid onto the processor grid. 

Our original implementation of such a method~\cite{acres}
demonstrated the feasibility of performing calculations within this 
framework. 
The data structures and operations involved in this method make it
easily parallelizable, particularly using high performance Fortran (HPF). 
In the present paper
we discuss several algorithmic issues that 
enhance the performance of the method and their 
implementation using HPF; we refer to the new implementation as HARES
for HPF-Adaptive-grid Real-space Electronic Structure.
The paper is organized as follows: 
In section~\ref{sec:theory_frame} we briefly review the theory
underlying the HARES  
method and present an analysis
of the computational effort involved in the various parts of the calculation. 
In section~\ref{sec:alg_enhance} we discuss the recent algorithmic 
enhancements and their implementation.
In section~\ref{sec:applications}
we illustrate the efficacy of these algorithmic enhancements 
through several applications of HARES to interesting systems.
These include: 
(a) a few simple elemental 
crystals and a few molecules composed 
of atoms in the first row of the periodic table which typically 
present a computational challenge to plane-wave (PW) methods; 
(b) blue bronze, a quasi one-dimensional
conductor; and (c) a zeolite, that is, a complex structure 
composed of Si-O tetrahedra and large pores, which
represents a molecular sieve. 
Section~\ref{sec:summary} contains our conclusions.

\section{Theoretical Framework}
\label{sec:theory_frame}

\subsection{Density Functional Theory
\label{DFT}}

The problem of finding the quantum mechanical ground state of electrons in
solids is a many body problem which, at present, 
can be solved only approximately.
The computational framework of choice for a wide range of problems 
involving a system of ions and interacting electrons is 
DFT \cite{dft}.
The central theorem of DFT, proven by Hohenberg and Kohn, 
states that the ground
state energy of an electronic system is a unique functional 
of its charge density 
$\rho(\rv)$ and is an extremum (a minimum) with respect 
to variations in the charge density. 
Kohn and Sham \cite{dft} expressed the charge density 
in terms of single particle 
wavefunctions $\psi_{\alpha}(\rv)$ 
(referred to as Kohn-Sham orbitals) and occupation numbers $f_{\alpha}$
\begin{equation}
\rho(\rv)=\sum_{\alpha} f_{\alpha} |\psi_{\alpha}(\rv)|^2.
\nonumber
\end{equation}
The ground state energy functional is then given by 
\begin{equation}
\begin{split}
\sum_{\alpha} f_{\alpha} \int \psi_{\alpha}^{*}(\rv) 
\left[-\frac{1}{2} \nabla^2 + 
V_{\rm ext}(\rv)\right] 
\psi_{\alpha} (\rv) d\rv + \\
E_{\rm\scriptscriptstyle H} [ \rho(\rv) ] + E_{\rm\scriptscriptstyle XC} [\rho(\rv) ],
\label{DFT_func}
\end{split}
\end{equation}
where $V_{\rm ext}(\rv)$ is the external potential experienced by 
the electrons due to the presence of the ions, $E_{\rm\scriptscriptstyle H}$ is the electrostatic (also 
known as Hartree) energy due to Coulomb repulsion of electrons 
and $E_{\rm\scriptscriptstyle XC}$ is the exchange-correlation (XC) contribution, which embodies the
many-body properties of the interacting electron system. 
A variational argument in terms of the single-particle states 
$\psi_{\alpha}(\rv)$ leads to a 
set of single-particle equations for fictitious non-interacting particles  
that produce the same density as the real electrons:  
\begin{equation}
\left[ -\frac{1}{2} \nabla^2 + V_{\rm eff}(\rho(\rv),\rv) \right] \psi_{\alpha}(\rv) =
 \epsilon_{\alpha}\psi_{\alpha}(\rv).
\label{DFT_eqns}
\end{equation}
The effective potential $V_{\rm eff}$ in these single-particle equations is:
\begin{equation}
 V_{\rm eff}(\rho(\rv),\rv) = V_{\rm ext}(\rv) 
+ V_{\rm\scriptscriptstyle H}[ \rho(\rv) ]
+ V_{\rm xc}[\rho(\rv)]
\end{equation}
where $V_{\rm\scriptscriptstyle H}$ is the electrostatic potential due to 
Coulomb repulsion between electrons (known as the Hartree potential)
and $V_{\rm\scriptscriptstyle XC}=\delta E_{\rm\scriptscriptstyle XC} / \delta\rho(\rv)$ 
is the exchange correlation potential. The system of 
Eqs.~(\ref{DFT_eqns}), referred to as Kohn-Sham equations, 
represents a set of nonlinear coupled equations 
due to the dependence of $V_{\rm\scriptscriptstyle H}$ and $V_{\rm\scriptscriptstyle XC}$ on the density 
(and hence the wave functions
$\psi_{\alpha}$); these equations are solved iteratively, 
beginning with a guess for the $\psi_{\alpha}$'s, 
until self-consistency is achieved.

The only significant approximation in this set of equations is the 
form of $E_{\rm\scriptscriptstyle XC}[\rho(\rv)]$, which is not analytically known.
The standard choices involve expressions that depend locally on $\rho(\rv)$
(known as the Local Density Approximation --- LDA), or involve both 
$\rho(\rv)$ and its gradients (known as the Generalized Gradient 
Approximation --- GGA).  Such expressions have been derived from analyzing 
the behavior of the uniform or non-uniform electron gas in certain 
limits, or by fitting the results of accurate calculations 
based on quantum Monte Carlo techniques for sampling the many-body 
wavefunction; they work well in reproducing the energetics of a wide variety 
of ground state structures of 
extended (crystalline) or finite (cluster or molecular) systems. 
We discuss next the basic features of the LDA and GGA approaches and
their limitations and capabilities. 

In the LDA, 
the exchange-correlation functional is expressed as:
\begin{equation}
E_{\rm\scriptscriptstyle XC}^{\rm\scriptscriptstyle LDA}[\rho({\rv})]
= \int \rho({\rv}) 
\epsilon_{\rm\scriptscriptstyle XC}^{\rm 0}(\rho({\rv}))  {\mathrm d}^3\rv.
\label{eq:XC-lda}
\end{equation}
when the number of spin-up and spin-down states in the system 
are equal (we refer to this as the ``spin compensated'' case).  
In this expression $\epsilon_{\rm\scriptscriptstyle XC}^{\rm
0}(\rho)$  is the exchange correlation energy of the {\em
uniform} electron gas of density $\rho$, which can be obtained by more
elaborate computational methods like quantum Monte Carlo, or even
analytically in certain limits.
The ground state of a number of physical systems, such as atoms,
molecules, and magnetic crystals can exhibit nonzero
spin-polarization.
In the above description of DFT, the single particle states label $\alpha$ 
can be extended to 
include the spin (up or down) quantum number, and 
the energy functional can be 
readily generalized to take into account the spin-polarized
electrons. The spin-polarized version of the  
exchange-correlation energy functional in the so called Local Spin
Density Approximation (LSDA) is written as
\begin{equation}
E_{\rm\scriptscriptstyle XC}^{\rm\scriptscriptstyle LSDA}
[\rho_\uparrow({\rv}),\rho_\downarrow({\rv})] = 
\int \rho({\rv}) \epsilon_{\rm\scriptscriptstyle XC}^{\rm 0}
(\rho_\uparrow({\rv}),\rho_\downarrow({\rv})) {\mathrm d}^3\rv,
\label{eq:Eks-spp}
\end{equation}
where $\rho_{\uparrow}(\rv)$ and $\rho_{\downarrow}(\rv)$ 
are the electronic densities of spin-up
and spin-down electrons, in terms of which the total electronic 
density $\rho(\rv)$ is: 
\begin{equation}
\rho(\rv) = \rho_{\uparrow}(\rv) + \rho_{\downarrow}(\rv).
\nonumber
\end{equation}

Among the various proposed XC functionals $\epsilon_{\rm\scriptscriptstyle XC}^{\rm 0}(\rho)$, 
in our approach to the DFT method  
we have implemented two different parametrizations of the Ceperley-Alder (CA) 
functional \cite{CA} 
for both spin-polarized and spin-compensated systems:
the first is the Perdew-Zunger \cite{PZ-CA}  parametrization (PZ-CA)
and the second is the Perdew-Wang \cite{PW-CA} parametrization (PW-CA).
The PW-CA functional uses a more accurate spin interpolation formula for
the correlation, proposed by Vosko, Wilk, and Nusair \cite{VWN},
which is based on 
the random-phase approximation; 
the PZ-CA functional uses the von Barth-Hedin \cite{BH}
spin-dependence for the correlation which is correct for the 
exchange part of the functional.

Although the L(S)DA has proved successful in a variety of chemical
and physical applications, it suffers certain well known deficiencies.
Among those, the most serious are:
\renewcommand{\labelenumi}{(\roman{enumi})}
\begin{enumerate}
\item The tendency to produce more bonding in solids than is observed 
      experimentally; manifestations of this tendency include 
      the underestimate of the lattice constant or bond length and 
      the overestimate of the cohesive energy and the bulk modulus.
\item Poor representation of activation energies which are related to 
      chemical reactions or transitions between structures.
\item Incorrect relative stability of different magnetic phases 
      for some magnetic materials.
\end{enumerate}
In order to correct these deficiencies, 
expressions for the XC functional which go 
beyond the density and include gradients of the density have been devised. 
In the so-called 
generalized gradient approximation (GGA), the XC energy functional  
is expressed as follows: 
\begin{equation}
\begin{split}
E_{\rm\scriptscriptstyle XC}^{\rm\scriptscriptstyle GGA}
[\rho_\uparrow({\rv}),\rho_\downarrow({\rv})] =\\
\int \rho({\rv}) 
\epsilon_{\rm\scriptscriptstyle XC}^{\rm\scriptscriptstyle GGA}
(\rho_\uparrow({\rv}),\rho_\downarrow({\rv}),
\nabla\rho_\uparrow({\rv}),\nabla\rho_\downarrow({\rv})) {\mathrm d}^3\rv.
\label{eq:Eks-gga}
\end{split}
\end{equation}
Applications of GGA to real materials show a tendency 
to over-correct the deficiencies of L(S)DA.
For instance, the lattice constants of 
common crystalline solids tend to be overestimated within GGA calculations,
while bulk moduli are underestimated~\cite{juan.1993.agc,juan.1995.ugg}.  
We have implemented in HARES 
the recently developed parameter-free GGA functionals 
(PW91 \cite{pw91,perdew96} and PBE96 \cite{perdew96,pbe96,perdew98}). 
Present capabilities include the use of the GGA functional in
two different modes : (a) fully self-consistent GGA calculations and 
(b) \textit{a posteriori} correction of the total energy with the 
perturbative GGA XC correction applied at the end of a 
self-consistent LDA calculation as:
\begin{equation}
\Delta E_{\rm total}
 = E_{\rm\scriptscriptstyle XC}^{\rm\scriptscriptstyle GGA}
[\rho^{\rm\scriptscriptstyle LDA}]
 - E_{\rm\scriptscriptstyle XC}^{\rm\scriptscriptstyle LDA}
[\rho^{\rm\scriptscriptstyle LDA}].
\label{eq:post-LDA}
\end{equation}
It is important to point out that the core--valence XC interaction is 
significantly different between LDA and GGA as was noted by 
Fuchs \textit{et al}. \cite{fuchs98} Therefore, in order to ensure
the reliability of the GGA results, it is necessary to perform either
the fully self-consistent GGA calculation using 
GGA-constructed pseudopotentials (referred to as mode (a) above) 
or the \textit{a posteriori} GGA correction after the self-consistent
LDA calculation using the LDA-constructed pseudopotentials 
with the partial-core electron density (referred to as mode (b) above).

For the spin-polarized systems, 
we consider two different modes of the computation:
(1) The conventional unconstrained calculation,  
where the total electron density and the magnetic moment are determined
simultaneously and self-consistently; and (2) the fixed spin moment (FSM) 
method \cite{FSM,moruzzi86,singh}, which constrains the magnetic moment to
be constant, but allows the possibility of different Fermi energies
for the spin-up and the spin-down electron densities.
The latter method has certain advantages: 
A series of FSM calculations with different magnetic moments provide
the total energy as a function of the magnetic moment, yielding 
detailed information about the magnetic phase. 
In addition, the FSM calculations rapidly achieve self-consistency
and are numerically more stable compared with the unconstrained
calculations.  

\subsection{Computational Approach
\label{comput_appr}}

There exist a variety of methods for solving the 
set of single-particle equations derived from DFT, Eqs.~(\ref{DFT_eqns}).
In the broadest classification, these methods fall into two categories,  
depending 
on how they describe the single-particle wavefunctions and charge density:
Methods in the first category use explicit basis sets to represent the
wavefunctions and charge density, while those in the second category 
use finite, discrete grids (or meshes) of points on which these functions 
are represented.\cite{jrc94,briggs95,baroni92,iyer95,hoshi95,gygi95}
A standard approach of the first type employs a PW basis, which
is a natural basis for periodic systems.\cite{ihm,jrc96}  
The plane waves needed in the expansion 
are determined by the reciprocal lattice of the 
crystal while the number of plane waves included
in the basis is determined by the highest kinetic energy, a parameter
referred to as the ``energy cutoff''.  
HARES falls in the second category of methods, as it employs 
a discrete mesh for the calculation. 
An important difference in the
two types of methods is that 
in the former all operators have a unique representation once the
basis set is chosen, whereas in the latter 
operators involving differentials have many
possible representations with different order of approximation. 
In this sense, the latter type of methods 
involve an additional degree of approximation. Both types of methods 
map the Kohn-Sham problem onto a matrix eigenvalue problem, 
denoted by $H_{\rm\scriptscriptstyle KS}$. 
One of the desirable features of grid-based methods is to produce 
a sparse matrix $H_{\rm\scriptscriptstyle KS}$; this makes it
possible to employ iterative algorithms for its solution. 

\subsubsection{Adaptive Coordinate Transformation
\label{coord_trans}}

HARES uses a uniform grid in curvilinear space which is analytically 
mapped onto a grid in real space with resolution (grid-spacing)
adapted to natural inhomogeneities in the problem. 
With the use of the adaptive grid, one can use HARES for both
all-electron and pseudopotential calculations.  However, use of
pseudopotentials proves effective in most practical calculations.
In the Kohn-Sham problem, inhomogeneities
arise fundamentally from $V_{\rm ext}(\rv)$, 
which is the potential that each 
electron experiences due to the presence of the nuclei. 
The Cartesian coordinates $x^i(\xi^{\alpha}; P^m)$
depend on the curvilinear coordinates $\xi^{\alpha}$ and 
a set of parameters $P^m$ that allows tuning of the coordinate 
representation to a particular physical problem. The Jacobian of the
transformation is 
\begin{equation}
J^i_{\alpha}({\bf  \xi};P)=\partial x^i/\partial \xi^{\alpha}
\end{equation}
and describes how derivatives transform between the coordinate systems;     
its determinant $|J|=\det J^i_{\alpha}$ is a measure of how 
the volume element is changed by the coordinate transformation. 
The metric giving the elemental length associated with infinitesimal
displacement is given by 
\begin{equation}
g^{\alpha \beta}=(J^{-1})_i^\alpha \delta^{ij} (J^{-1})_j^\beta.
\end{equation}
Details of the coordinate transformation can be 
found in Ref.~\onlinecite{acres}.

The gist of the transformation is to enhance spatial 
resolution in the region where
it is desirable to increase the accuracy of the finite-difference derivatives
and the representation of charge density inhomogeneities. 
The equivalent enhancement of resolution in the PW approach is
the increase of the energy cutoff.  The connection between the 
effective energy cutoff and the local resolution of the HARES grid is given 
by the factor $|J|^{-2/3}$.
The differential equation
of the Kohn-Sham problem in the adaptive grid representation 
becomes a finite matrix eigenvalue problem, 
with only the kinetic
energy term (the Laplacian in the single-particle 
equations) having off-diagonal elements. 

The uniform mesh in $\xi-$coordinates is subsequently 
broken into blocks that are distributed over a number of processors on a
parallel computer architecture. The wavefunctions, potentials, 
and charge density are represented on this
mesh allowing for balanced distribution on processors. 
In the iterative solution of the
eigenvalue problem, an operation that is performed frequently
during the calculation  
is the product of the Hamiltonian matrix with 
a vector representing a single-particle wavefunction.
In parallel execution of this operation, it is the  kinetic energy term 
(the Laplacian) with off-diagonal elements that requires most  
of the communication and makes the solution of the 
eigenvalue problem nontrivial.

\subsubsection{Boundary Conditions
\label{bound_cond}}

\begin{figure}
\psfile{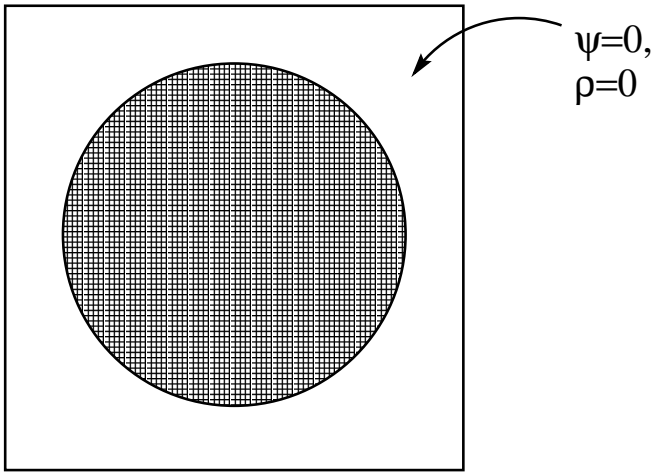}{1.0}
\caption{Cell used in a calculation with open boundary conditions.}
\label{fig:OBC}
\end{figure}

In DFT calculations based on a real-space grid, 
boundary conditions enter in the 
way the Laplacian is applied to a function.  
There are only two aspects of a calculation
where this is relevant: 
(i) the kinetic energy operator, that is, 
    the Laplacian acting on wavefunctions and 
(ii) the calculation of the electrostatic potential, 
     obtained by solving the Poisson equation, that is,  
     the Laplacian acting on the potential. 
Since the calculation of the Laplacian (represented as a finite 
difference) of a function at a given grid point uses values of 
the function at adjacent grid points, 
imposition of the boundary conditions requires knowledge of
the function at a few grid points outside the boundary.
For calculations on infinite crystalline solids, 
we use periodic boundary conditions (PBC)
demanding that the function is periodic in space with the period of
a unit cell that models a physical system.
Thus, application of the Laplacian at any point in the unit cell involves
values of the function inside the unit cell making implementation of
boundary conditions for this case straight-forward.
For calculations on finite systems 
(atoms, molecules or clusters),
we use open boundary conditions (OBC). 
In this case, the treatment of boundary conditions
is more intricate since the grid points adjacent to the ones on
the boundary fall both
inside and outside the region containing the system. 
In Fig.~\ref{fig:OBC}, we show
how this is treated in HARES. 
We choose a rectangular box with a spherical region
inside, the interior of which is large enough to contain the physical system. 
The thickness of the buffer region
surrounding this sphere
depends on the order of the finite-difference Laplacian, 
and is equal to this order times
the grid-spacing.
Thus, the distance between the sphere and the box vanishes in the
continuum limit and the ratio of the volumes of the sphere
to the box becomes $\pi/6$.
The wavefunctions and
charge density vanish outside the spherical region; 
this takes care of the boundary
conditions in the kinetic energy part. 
There is one more term that deserves special attention 
in handling OBC: it is 
the electrostatic potential which has long range,
and therefore cannot be assumed to be zero in the buffer region 
outside the spherical region.
We obtain values for the electrostatic potential in the buffer region 
through a multipole expansion
up to order 4, using the density inside the spherical region;
we use the values at those points as the boundary
conditions for $V_{\rm\scriptscriptstyle H}$ in the solution of 
the Poisson equation.

\subsubsection{Scaling of Computational Effort
\label{scaling}}

It is useful at this point to 
analyze the computational effort involved in
various aspects of a DFT calculation using HARES and compare it to that
performed using the PW method. 

In the PW method, there is one grid in real (direct) space and
another in reciprocal space. The wavefunctions are stored in reciprocal space
on part of the grid inside a spherical region with diameter 
equal to half of the side
of the full grid. These are transformed into real space using the fast Fourier
transform (FFT) whenever necessary. 
The calculations of the charge density, the local ionic potential, and the  
exchange correlation energy are carried out in real space, whereas 
the calculation of the 
kinetic energy and the Hartree energy are carried out in reciprocal space. 
The calculation of the energy related to the nonlocal
pseudopotential can be done on either grid. For small system sizes, 
the most time-consuming part is often that of
performing the  FFTs, which scales as $O(N\log N)$.
The number of FFTs scales linearly with system size giving an overall 
$O(N^2\log N)$ scaling for the entire calculation. 
For large system sizes, the orthonormalization of wavefunctions, 
or equivalent constraints imposed during minimization of the energy functional, 
dominate the computational time.  These operations scale 
as $O(N^3)$. Calculation of the contribution from the nonlocal 
pseudopotentials normally scales as $O(N^3)$, 
but can in principle be improved
to $O(N^2)$ scaling by exploiting the short-range character 
of the potentials in real space.

\begin{table}
\caption{Scaling of computational effort with system size 
         in HARES and a PW method.}
\label{tab:scaling}
\begin{center}
\begin{tabular*}{\columnwidth}{@{\extracolsep{\fill}}l|c|c}
\toprule
   Calculation of     &  HARES    &   Plane Wave Method  \\
   \colrule
   Charge density     & $O(N^2)$  &   $O(N^2 \log N)$    \\
   Kinetic energy     & $O(N^2)$  &   $O(N^2) $          \\
   Local potential    & $O(N^2)$  &   $ O(N^2 \log N)$   \\
   Nonlocal potential & $O(N^3)$  &   $O(N^3)$           \\
   Hartree energy     & $O(N)  $  &   $ O(N) $           \\
   XC functional     & $O(N)  $  &   $ O(N) $           \\ 
   Orthogonalization  & $O(N^3)$  &   $ O(N^3)$         \\  
\botrule
\end{tabular*}
\end{center}
\end{table}
In HARES, the wavefunctions are stored on the full grid in real space and all
operations are performed on the same grid, eliminating the need
for FFTs. The calculation of the kinetic energy is carried out 
using finite-difference
formulae for derivatives on a grid in real space. 
The Hartree energy and the long-range 
electrostatic potential due to periodic charge density 
are computed by solving the
Poisson equation, which scales as $O(N)$. 
For large system sizes, orthonormalization of the wavefunctions 
dominates the computational time, which then
scales as $O(N^3)$. The treatment of the  
nonlocal pseudopotentials also scales as $O(N^3)$, 
unless their short-range character is exploited. 
We  summarize the comparison of scaling between HARES and a PW method 
in Table~\ref{tab:scaling}.

The current parallel implementation of HARES is 
in high performance Fortran (HPF), which 
involves single instruction multiple data (SIMD) coding. 
Since the wavefunctions and charge density are stored in real space 
and distributed across processors, communication between processors 
is necessary in calculating:
\renewcommand{\labelenumi}{(\alph{enumi})}
\begin{enumerate}
\item the finite-difference derivatives using the CSHIFT operation, 
        which cyclically shifts the data in an array on a grid 
        along the specified direction; and
\item the inner product of two functions using the the SUM operation.
\end{enumerate}
The scaling of inter-processor communication with system size
is presented in Table~\ref{tab:communication}. 
For large enough system size, the SUM operations dominate 
the communications in a parallel calculation. 
\begin{table}
\caption{Scaling of communication between processors (assumed to be 
         on a cubic array) used in HARES calculation.}
\label{tab:communication}
\begin{center}
\begin{tabular*}{\columnwidth}{@{\extracolsep{\fill}}l|c|c} 
\toprule
   HPF        &  Communication   &  Number of  \\
   operation  &   per operation  &  operations \\
   \colrule
   CSHIFT     & $O(N^{2/3})$     & $O(N)$      \\
   SUM        & $O(N^0)$         & $O(N^2)$     \\
\botrule
\end{tabular*}
\end{center}
\end{table}

\section{Algorithmic Enhancements and Implementation}
\label{sec:alg_enhance}

\subsection{Non-orthogonal Unit Cell}
\label{unit_cell}

Finite-difference formulae for derivatives of functions represented on
a grid with finite spacing are designed to achieve high accuracy and
are reasonably accurate for a polynomial function up to certain order. 
These formulae are typically derived
for functions of one variable on grids of uniform spacing. Their 
generalization to higher dimensions is trivial through direct product
if the grid is orthogonal in the various dimensions. 
For periodic crystals with non-orthogonal unit cell,
it is often not possible to design an orthogonal grid with the same periodicity 
in all directions; in this case 
the implementation of finite-difference formulae is not trivial.

The coordinate transformation employed in HARES provides a very simple
method to treat non-orthogonal unit cells and grids. In this case, the
transformation is uniform throughout the unit cell and maps an orthogonal
unit cell in $\xi-$space onto a non-orthogonal unit cell in $x-$space. If
$F$ is a matrix that gives the deformation of the orthogonal unit cell into
the one under study 
({\it i.e.} its columns are the non-orthogonal unit cell vectors),
one can always obtain a transformation that is symmetric by filtering  
out the rotational part of $F$ as follows: first obtain an auxiliary
matrix $M$ defined as
$M=F\cdot F^T$ and then diagonalize $M$ to obtain a diagonal matrix $D$. 
The Jacobian for 
a rotation-free transformation is then given by 
\begin{equation}
J=T^{-1} \cdot D^{\frac{1}{2}} \cdot T,
\nonumber
\end{equation}
where $T$ is a matrix that diagonalizes $M$: $D = T \cdot M \cdot T^{-1}$. 
Once the mapping onto an orthogonal $\xi-$grid is obtained, 
the derivatives, length and 
volume elements can be obtained using the formalism described in 
section \ref{coord_trans}.

\subsection{Preconditioned Conjugate Gradients Solver}
\label{precon_cg}

The dominant part of a DFT calculation often consists of 
solving an eigenvalue problem, that is, obtaining the lowest few 
(compared with the full spectrum) eigenvalues and eigenvectors 
of a very large matrix. 
In the case of PW basis, the size of the matrix is determined by the 
number of PW components included in the basis. 
In the case of HARES, the size of the matrix is determined by the number 
of points that constitute the real space grid. 
For small enough matrices the standard techniques of linear algebra 
can be employed, which give the 
exact (within the numerical accuracy of the algorithm) eigenvalues 
and eigenvectors of the matrix.  When the size of the matrix is large, 
the conventional methods are not practical and the only alternative  
is to employ iterative approaches which approximate the eigenvalues and 
eigenvectors in successively improving steps.
We considered two iterative algorithms for the diagonalization task 
in HARES: 
\renewcommand{\labelenumi}{(\alph{enumi})}
\begin{enumerate}
\item An Inverse Iteration (II) algorithm with multigrid 
      preconditioning \cite{norm_thesis};
\item A Conjugate Gradient (CG) algorithm \cite{conj} with suitable 
      preconditioning in real space. 
\end{enumerate}
The implementation of the former has been presented earlier \cite{acres}
and we want to focus on the CG algorithm in this subsection.

\begin{figure}
\psfile{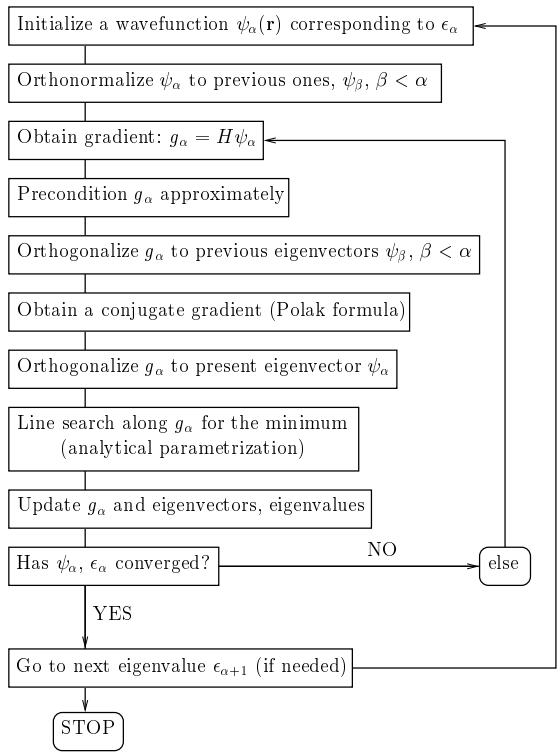}{0.85}
\caption{Flowchart of the conjugate-gradient algorithm for the eigenvalue problem.}
\label{fig:CG}
\end{figure}

In Fig.~\ref{fig:CG}, we present a flowchart of the CG algorithm. 
It is similar to the one in Ref.~\onlinecite{conj}, presented for a
PW basis. In real space, most steps in the algorithm 
remain unchanged except for the preconditioning. 
The main idea in preconditioning is to filter out high Fourier
components in the wavefunctions and the charge density. 
We achieve this through multiple application of a coarsening
transformation.  For example, the coarsening applied to the charge density
gives: 
\begin{equation}
\begin{split}
\rho(k,l,m) \rightarrow 
\frac{1}{2} \rho(k,l,m) + \\
 \frac{1}{12} [\rho(k \pm 1,l,m)+ 
\rho(k,l \pm 1,m)+\rho(k,l,m \pm 1)],
\end{split}
\end{equation}
where $k,l,m$ are indices of the grid points at which 
the charge density $\rho$ is calculated.
We find that application 
of this transformation twice on the function under consideration 
results in adequate preconditioning.  Better preconditioning is
possible in principle for selected cases but may not be worth the
extra effort required; this
method provides a preconditioner that works reasonably well in all
cases we have considered.

Another aspect of a DFT calculation is that the eigenvalue 
problem needs to be solved repeatedly 
while updating the charge density to achieve self-consistency between the 
wavefunctions and the corresponding effective potential, 
which depends on the density. 
Specifically, a self-consistent DFT calculation starts 
with an initial guess for the density and subsequently 
follows an iterative procedure of alternating steps of 
diagonalization and improved estimate for the charge density. 
Each iteration $i$ starts with a charge density $\rho_{in}^i$
and obtains the part of the eigenspectrum of $H_{\rm\scriptscriptstyle KS}$
that corresponds to occupied single-particle states and some
low-lying unoccupied states, 
which depends on $\rho_{in}^i$ through the effective potential. 
At the end of an iteration the eigenfunctions are used to obtain 
an output charge density $\rho_{out}^i$. An improved 
estimate for the charge density is
obtained from $\rho_{in}^i$ and $\rho_{out}^i$; 
this is discussed in detail in the following subsection.

We observe that diagonalization with full convergence,  
that is, with the tolerance for the difference between 
the $\rho_{in}^i$ and $\rho_{out}^i$ set to be very low, 
can be computationally demanding and depends on the initial guess 
for the density. Since the charge density at the
initial stage of self-consistency is far away from the 
self-consistent one, accurate diagonalization of the Hamiltonian 
matrix at this stage is not worthwhile. 
Accordingly, we limit the number of CG steps for diagonalization 
at the initial stages of the self-consistency loop to 
relatively few --- we found that two CG iterations at 
this stage yield optimal efficiency. 
Accurate diagonalization is achieved in the course of
self-consistency as the initial guess for the
eigensolver is steadily improved.
We have found that the performance of the II and CG
algorithms for convergence to self-consistency is comparable. 
The differences are rather small and system-dependent. 
The advantage of the CG algorithm is that all operations 
take place on the same grid, whereas in the II algorithm 
multi-grid preconditioning requires the mesh sizes to be 
a power of two.

\subsection{Charge Density Mixing \label{mixing}}

We return now to the way in which the charge density is updated 
at the end of each iteration. 
In general, the new charge density at the end of step $i$ 
can be constructed from the charge densities of previous steps: 
\begin{equation}
\rho_{new}^i = 
\sum_{j=i-d}^i ( \kappa_j^{in} \rho_{in}^j + \kappa_j^{out} \rho_{out}^j )=\rho_{in}^{i+1},
\end{equation}
where $\kappa_j$'s are mixing coefficients and 
$d$ is called the ``depth" of the mixing
procedure, {\it i.e.} the number of previous iterations 
used in the improved estimate of the density.
Various mixing schemes are available and have been 
discussed in Ref.~\onlinecite{kresse}.

In our work, we added another feature to mixing: 
we optimize the strength of mixing, that is, the values of the 
parameters $\kappa_j$, as a function of iteration. 
This feature can be used along with most of the mixing schemes
employed in the literature. To illustrate the basic idea, we take a 
simple realization of a mixing scheme:
\begin{equation}
\rho_{new}^i = \rho_{in}^i + \kappa (\rho_{out}^i-\rho_{in}^i).
\end{equation}
$\kappa=1$ is an extreme case where no knowledge of the input density is used,
and $\kappa=0$ corresponds to the opposite
extreme where the density is not updated at all.
For $\kappa=1$, there can be oscillations between 
input and output densities corresponding
to underdamped mixing. On the other hand, for small $\kappa$ 
the oscillations are
overdamped resulting in slow update of the density and hence 
the approach to self-consistency.
We have devised a way to achieve the optimal \textit{critical damping}
in the mixing procedure.
In our scheme, we calculate the root mean square change in density
$\delta \rho^i$ at each iteration $i$:
\begin{equation}
\delta \rho^i = \left( \frac{1}{\Omega} \int |\rho^i_{out}-\rho^i_{in}|^2 {\mathrm d}^3 \rv \right)^{\frac{1}{2}},
\nonumber
\end{equation}
where $\Omega$ is the unit cell volume, 
and define the rate of self-consistency as 
\begin{equation}
R_i\equiv\frac{\partial \log (\delta \rho^i)}{\partial t},
\nonumber
\end{equation}
where $t$ is a fictitious time associated with iterations. 
$R_i$ is a rough measure of how well a calculation is evolving 
toward a self-consistent solution. 
We monitor both the rate and the mixing coefficient at each iteration 
and can estimate $\lambda \equiv \partial R/\partial \kappa$.
If $R_i$ is too small ($<0.2$), 
the approach to self-consistency is too slow. 
In that case, the mixing coefficient is increased or decreased, 
depending on the sign of $\lambda$, by an amount $\Delta \kappa$, 
--- 
a positive $\lambda$ corresponds to smaller $\kappa$ and 
a negative $\lambda$ to larger $\kappa$ 
compared with the optimal value of this parameter. 
This has the effect of keeping the strength of mixing
$\kappa$ near a value where $R$ is optimal \textit{i.e.}, 
in the neighborhood of critical damping. 
The magnitude of change $\Delta \kappa$ is reduced over time 
to make sure that it converges to its optimal value.

\subsection{Nonlocal Pseudopotential Projector \label{psp_project}}

The nonlocal part of the pseudopotentials $V_{NL}(\rv,\rv^{\prime})$ is used in a separable 
form \cite{kleinman,gonze} to facilitate fast calculation of its product 
with the wavefunction $\psi$:
\begin{equation}
\langle \rv | V_{NL} | \psi \rangle
 = \sum_{\alpha, l, m} 
\frac{ \langle \rv|\phi_{lm}^{\alpha}\rangle\langle \phi_{lm}^{\alpha}|\psi\rangle}
{\eta_{lm}^{\alpha} }, 
\nonumber
\end{equation}
where $\alpha$ is an atomic index, $l$ and $m$ are angular quantum indices, and
$\eta_{lm}^{\alpha}$ are constants related to the normalization of the nonlocal
projectors $\phi_{lm}^{\alpha}$.
Straightforward application of $V_{NL}(\rv,\rv^{\prime})$ on 
the wavefunctions scales as $O(N^3)$, since
there are $N_e$ wavefunctions on $N$ grid points, and $N_a$ atoms 
(the number of grid points $N$ is proportional to the system size
measured by either $N_a$ or $N_e$, which also scale with each other). 
This computation can
be accelerated significantly, by noticing that 
$\langle \rv|\phi_{lm}^{\alpha}\rangle$ is a localized
function centered on atom $\alpha$. 
Thus, a calculation of $\langle \phi_{lm}^{\alpha}|\psi\rangle$
in real space involves only the grid points near atom 
$\alpha$ making it an $O(N^0)$ 
computation, which gives $O(N^2)$ for the overall calculation 
involving the non-local pseudopotential.
We use a filtered pseudopotential
approach~\cite{bernholc:pseudopotentials} with a filter 
that is smooth in $k$-space (as opposed to the theta-function used in
plane-wave approaches).  This minimizes the errors in the evaluation
of $\langle \phi_{lm}^{\alpha}|\psi\rangle$ in addition to using the
adaptive grid.  
The parallel implementation of such a calculation is not trivial, 
since it involves only those 
processors which store the grid points near the given atom $\alpha$. 
To address this issue, we represent 
the atomic projectors $\phi_{lm}^{\alpha}(\rv)$ in terms of the packed
projectors $\chi_{lm}^{j}(\rv)$ shown schematically as follows:
\begin{equation}
\begin{split}
\left\{ \phi_{lm}^{\alpha}(\rv) | \alpha = 1, \ldots, N \right\}
\longrightarrow
\\
\left\{ \chi_{lm}^{j}(\rv), \beta_{lm}^{j}(\rv) | j=1, \ldots, M_d  \right\}
\label{local_proj}
\end{split}
\end{equation}
where $\chi_{lm}^{j} (\rv)$ is defined as the projector of the $j$-th
largest magnitude at a given grid point $\rv$
(e.g.\ $\phi_{lm}^{\gamma}(\rv)$), and $\beta_{lm}^{j}(\rv)$ is the
index of the atom from which $\chi_{lm}^{j} (\rv)$ was generated (in the
case of the above example $\beta_{lm}^{j}(\rv) = \gamma$).
We only keep a number of important pseudoprojectors $M_{\rm d}$,
(thus letting $j$ vary from 1 to $M_{\rm d}$), 
which we call the depth of the packed projectors.
We find that $M_{\rm d} = 3$ is typically sufficient;
for example, this is exact if the nonlocal projectors of at most three
atoms are nonzero at any point $\rv$.
This changes the scaling of memory requirements for the nonlocal potential
from $O(N^2)$ to $O(N)$.  
With this choice of packing, the expression for  the inner product becomes:
\begin{equation}
\langle \phi_{lm}^{\alpha}|\psi\rangle=
\sum_j \sum_\rv 
\delta_{\beta_{lm}^{j}(\rv), \alpha}  
\langle \chi_{lm}^{j}|\rv\rangle \langle \rv|\psi\rangle.
\nonumber
\end{equation}
This is readily evaluated using an 
EXTRINSIC subroutine call in HPF, which involves
execution of the whole routine on each processor, but on different data. 
Effectively, an inner product $\langle \phi_{lm}^{\alpha}|\psi\rangle$ with contribution
from only the grid points inside a sphere centered at atom 
$\alpha$ is calculated
by distributing the data with respect to $\alpha$ rather than grid points.

\subsection{Computation of Forces \label{forces}}

Forces on the atoms are calculated using the Hellman-Feynman theorem,
as is usual in DFT calculations:
\begin{equation}
\Fv_{\alpha} = -
\sum_i \left\langle \psi_i \left| \frac{\partial V_{\rm ext}}{\partial \Rv_\alpha} \right| \psi_i \right\rangle,
\nonumber
\end{equation}
where $\Rv_\alpha$ is the position of atom $\alpha$.
For reasons similar to those mentioned in the previous subsection, 
the contribution of the
nonlocal pseudopotential to atomic forces is computationally 
demanding and scales as 
$O(N^3)$. With packed projectors $\langle \rv|\chi_{lm}^{j}\rangle$, the
scaling of this computational cost
is improved to $O(N^2)$. 
This is a significant improvement for problems involving structural relaxation
of large systems.

The implementation of packed projectors in the calculation of forces 
deserves further elaboration. The force
calculation with nonlocal pseudopotentials involves 
both the projectors $\phi_{lm}^{\alpha}(\rv)$ and
their derivatives $\partial \phi_{lm}^{\alpha}/ \partial \Rv_{\alpha}$. 
Since the latter is needed
only during the calculation of forces, 
it does not need to be packed and stored but can be  
obtained at the time when it is needed. 
The inner products of $\phi_{lm}^{\alpha}$ with $\psi_i$ 
have to be computed for all atoms $\alpha$ at once 
at the beginning of a force calculation since they are packed.  
With these improvements, we 
obtain a factor of 7 speedup in the 
calculation of forces for systems containing about 30 atoms. 

Finally, we should note that as in any method that involves a
computational basis which changes with the positions of the atoms,
the adaptive grid generates fictitious forces referred to as 
Pulay forces.  We have implemented the correction related to the 
Pulay forces and found that it is not significant when the 
grid is refined to the point where it yields adequate accuracy.
Thus, for all practical purposes, with an accurate grid the Pulay  
correction to the forces can be neglected (see also Ref.~\onlinecite{acres}).  

\subsection{Geometry Optimization \label{sec:relax}}

    In modern \textit{ab initio} total energy calculations, 
one of the objectives is to obtain 
minimum energy geometries (corresponding to local minima
or, if possible, the global minimum
of the energy), where the Hellmann-Feynman 
force on each atom is zero or, more precisely, smaller in magnitude 
than a prescribed value (typically $\le 0.5$~mRy/a.u.).
The process in which the initial ionic geometry is 
sequentially updated to relax to a neighboring local minimum is referred to 
as the \textit{ionic relaxation}. Mathematically, the ionic relaxation
is a  
nonlinear optimization problem, which is 
a subject of vast interest in applied mathematics.

   Many optimization algorithms have been proposed so far,  
which can be broadly classified into three groups:
\renewcommand{\labelenumi}{(\roman{enumi})}
\begin{enumerate}
  \item algorithms which require only the evaluation of the function;
  \item algorithms requiring the function values and its gradients;
  \item algorithms requiring the function values and its first and 
        second derivatives (the gradients and the Hessians).
\end{enumerate}
In electronic structure calculations, 
the function to be optimized is the total energy and its gradients 
are the forces on the atoms. 
Since the forces are not too computationally demanding compared
to the self-consistency loop (especially after implementing
the improvement discussed
in the previous subsection), it is natural 
to use the methods of class (ii) for ionic relaxation.

   Among the gradient algorithms, the quasi-Newton (also referred to as 
variable metric) method is known to be most efficient \cite{brodlie77}. 
The inverse Hessian is approximated and updated at each iteration.
Suppose $|R_0\rangle$ is an initial estimate
of the minimizer of the total energy $E_{\rm total}$, $|g_0\rangle$ is 
the corresponding gradient, and $H_0$ is the initial guess 
for the inverse Hessian.
At the $n$-th relaxation step, the next approximate minimizer is given by
\begin{equation}
   |R_{n+1}\rangle = |R_n\rangle - \beta_n H_n |g_n\rangle
\end{equation}
where the step size $\beta_n$ is determined by the line 
search \cite{walsh75} (or the line minimization).
The inverse Hessian is updated by
\begin{equation}
   H_{n+1} = H_n + \Delta_n
\end{equation}
where $\Delta_n$ is the correction to $H_n$ and is determined by
requiring that it satisfies the quasi-Newton condition, 
\begin{equation}
   H_{n+1} | h_n \rangle = | d_n \rangle
\end{equation}
with $| h_n \rangle\equiv | g_{n+1}\rangle - | g_n\rangle$ and
$| d_n \rangle\equiv | R_{n+1}\rangle - | R_n\rangle$.

Among different update formulae for the inverse Hessian \cite{brodlie77}, 
we have implemented the initially-scaled Broyden-Fletcher-Goldfarb-Shanno 
(IS-BFGS) expression \cite{shanno78,chetty95}:
\begin{equation}
\begin{split}
 H_{n+1} = \tilde{H}_n
   - \frac{\tilde{H}_n|h_n\rangle\langle d_n|
          + |d_n\rangle\langle h_n|\tilde{H}_n}{\langle d_n | h_n\rangle}
\\
   + \biggl( 1
       + \frac{\langle h_n | \tilde{H}_n | h_n \rangle}{\langle d_n | h_n\rangle}
     \biggr) \frac{\langle d_n | d_n\rangle}{\langle d_n | h_n\rangle} 
\label{eq:IS-BFGS}
\end{split}
\end{equation}
where
$\tilde{H}_n = (\langle d_n | h_n \rangle / \langle h_n | H_n | h_n \rangle) H_n$,
when $n=0$ and $\tilde{H}_n = H_n$, otherwise.

Implemented in combination with the approximate 
line search algorithm \cite{fletcher70}, the IS-BFGS provides 
an efficient ionic relaxation tool which assures the convergence 
of the approximate inverse Hessian to the correct inverse Hessian 
and is numerically stable. 
As noticed in other works \cite{powell77,shanno78b}, 
restarting the update sequence [Eq.~(\ref{eq:IS-BFGS})] can be beneficiary 
in some cases, and we have used the following restart criterion:
$\langle d_n | g_{n+1}\rangle \leq 0.5 
\sqrt{\langle d_n | d_n\rangle\langle g_{n+1}| g_{n+1}\rangle}$.
This implies that the displacement of an atom is not too different
from the direction of the calculated force acting on it.

\subsection{Dual real-space grid calculations
\label{dual_grid}}

Typically, the representation of the charge density and 
local potentials in a DFT calculation needs twice as much 
spatial resolution as that of the wavefunctions. 
To exploit this aspect of electronic structure calculations, 
we have developed a version of HARES 
which employs two separate real-space grids --- 
a coarser one and a finer one. 
The wavefunctions are represented on the coarser grid and 
the charge density on the finer one with half the grid-spacing 
of the former. The finer grid corresponds to the FFT grid 
in a PW calculation.  This version of the code reduces
memory requirements substantially. The computational 
cost of the worst scaling part of the calculation 
(the wavefunction orthogonalization) is reduced by a factor of 8.

The transformation from the coarser to the finer grid is 
performed only when the charge density needs to be calculated from 
the wavefunctions. We use wavelet interpolants~\cite{arias} 
to achieve this. 
The Poisson equation is solved on the finer mesh to obtain the 
electrostatic potential. The exchange correlation potential 
and the local part of the pseudopotential are also calculated 
on the finer grid; these terms are convolutions in $k$-space and in
a PW method are calculated on the FFT grid.
 The kinetic energy operator 
and the nonlocal pseudopotential act on the wavefunctions 
directly on the coarser grid.
Naturally, this necessitates usage of a higher order Laplacian 
in the calculation of the kinetic energy, while the one 
with lower order is adequate in the solution of the Poisson equation.

To check the accuracy of the dual-grid code, we calculated the energy
difference between two configurations of the O$_2$ molecule and
compared it with the result of a PW calculation. Both methods were
used at different energy cutoffs, or equivalently, of spatial
resolution.  We found that the energy difference as a function of the
energy cutoff behaves the same way in both methods.  In fact, at a low
energy cutoff, the sign of the energy difference was inverted in both
calculations and the magnitude was within 6~\% of the correct value.

While the dual-grid approach to real-space electronic structure 
calculation enhances the performance significantly, we caution 
that the errors introduced by breaking of translational invariance,
a feature inherent in real-space grids (see Ref.~\onlinecite{acres}),
are larger than those in the single finer
grid calculations. 
This is due to the coarser grid used in the representation of wavefunctions 
and the higher order expression for the Laplacian on the coarser grid. 
Within the PW method, the same error enters
in calculating the XC potential on the real-space FFT mesh.
These errors enter into the calculation of 
forces and can be minimized if necessary by 
using Fourier-filtered pseudopotentials. 
We expect the use of the dual-grid approach 
to be very advantageous at the initial stages
of ionic relaxation of a system with large number of atoms, 
when the accuracy in forces or wavefunctions is not crucial since the 
system is presumably far from its optimal structure.

\subsection{Performance
\label{performance}}

\begin{table}
\caption{Comparison of performance of HARES and a PW method.}
\label{tab:acres-pw}
\begin{center}
\begin{tabular*}{\columnwidth}{@{\extracolsep{\fill}}l|c|c} 
\toprule
   System            &  CPU time (8 nodes)   & CPU time (1 node)  \\
                   &     HARES             &  CASTEP 2.1        \\
\colrule
   O$_2$             &    150 sec            &     1548 sec       \\
   Si$_{24}$O$_{48}$ &    7.5 hours          &     36 hours      \\ 
\botrule
\end{tabular*}
\end{center}
\end{table}

In Table~\ref{tab:acres-pw}, we present a comparison of the 
performance of HARES and that of a PW code, for DFT 
calculations on the O$_2$ molecule and a zeolite, Si$_{24}$O$_{48}$. 
We have selected for the comparison the academic version of
CASTEP~\cite{castep}, a PW package which uses all the standard methods
for such calculations and is freely available to academic researchers,
as is also the case for HARES.   We believe this provides the most
meanigful comparison of the performance of the different approaches,
for codes at equivalent levels of development and availability to the
academic community.  
It is clear that the performance of
HARES for the oxygen molecule is definitely better than the 
PW code.  In general, we find that HARES performs better 
for metallic systems. For the zeolite, which is an insulator, 
CASTEP uses a variational method for direct minimization
of the total energy, which is not applicable to the case of metals.
This makes the performance of CASTEP very good for such systems,
though the performance of HARES is not unacceptable (a factor of 1.66
slower than CASTEP).  This advantage of CASTEP over HARES is lost when
applied to metallic systems, in many of which the academic version of 
CASTEP available to us failed to converge in the self-constistency
loop.   We have been informed that in commercial versions of this
package the problematic convergence to self-consistency in metallic
systems has been solved and the performance has been
improved~\cite{castep-extra}. 
We point out that a comparison of {\it ab initio} packages, 
is meaningful only for methods that employ 
the same type of pseudopotentials.  
In the comparison discussed here,
both methods use 
conventional norm-conserving pseudopotentials. There exists 
another class of pseudopotentials, called ultra-soft 
pseudopotentials~\cite{dhv}, which reduce the size of the Hamiltonian
matrix substantially, making the calculations more efficient.
Codes that employ this class of pseudopotentials are naturally 
faster, whether they use a PW basis or a real-space grid.  
For instance, the VASP code 
based on the PW formulation uses ultra-soft pseudopotentials and
has proven quite effective\cite{vasp}; the commercial
version of CASTEP also uses these pseudopotentials~\cite{castep-extra}.
This class of pseudopotentials has not been yet implemented in HARES.
 
\section{Applications}
\label{sec:applications}

As a test of the accuracy and the efficiency of the algorithmic improvements 
discussed above, we offer a range of example applications of HARES. These 
include representative elemental crystals, some molecules, and a couple of 
rather complex materials --- blue molybdenum bronze and the TON zeolite. 
All the calculations were performed on a Silicon Graphics Origin 2000, 
using from 2 to 16 processors in parallel mode.

\subsection{Study of elemental solids \label{solids}}

The simplest test of the method is its application to elemental
crystalline solids.  We have calculated the basic structural and
electronic properties for representative elemental solids,
including alkali metals (Li, K), group II A metals (Be, Ca),
sp-electron metals (Al, Ga), d-electron non-magnetic metals (V, Cu,
Mo), d-electron magnetic metals (Fe, Ni), and semiconductors and
insulators (Si, C).  The properties of these solids are extracted from
total energy calculations for a given crystal structure, 
using the LDA and applying the {\em a posteriori} GGA
corrections.  We have used norm-conserving pseudopotentials from
Bachelet~\textit{et al}~\cite{bachelet.1982.pwh}
for V, and pseudopotentials
generated with the Troullier and Martins~\cite{troullier91} scheme for
the all the other elements.  
We perform the calculations as follows: we
choose a sufficiently dense grid of \textbf{k}-points in the
Monkhorst-Pack scheme~\cite{monkhorst.1976.spb} and fold it to the
irreducible part of the Brillouin zone by
applying the 
symmetry operations of the point group of the crystal including
inversion
which is always a symmetry operation in reciprocal space.  We
make sure the calculation is converged with respect to 
the real space grid spacing in the 
neighborhood of the anticipated equilibrium lattice constant, and keep the
grid spacing approximately constant
for a range of lattice constants up to about twice the equilibrium
value.  
We then fit the resulting 
energies to powers of $\Omega^{-2/3}$, where $\Omega$ is the
volume of the unit cell.  We thus obtain accurate values of
the equilibrium lattice constant and minimum total energy.  These values
are used to fit the two-parameter Universal Binding Energy
Relation~\cite{rose.1984.ufe}, which has a simple analytical form
from which the bulk modulus and the cohesive energy are obtained.
This procedure relies on the fact
that the total energy differences used to fit the Universal Binding
Energy Relation converge quicker than the total energy with respect to
the grid parameters.
For the nonmagnetic materials we use the non-spin-polarized code to
perform these calculations for reasons of computational efficiency.
The free atoms in most of the cases we calculated are
polarized, the extreme case being Molybdenum were all six valence electrons
have the same spin.  We thus add to the cohesive energy as calculated
previously the difference of a free atom calculation using the
spin-polarized and spin-average codes.  The free atom calculations can
be converged to 
the same, high degree of accuracy by use of the adaptive grid and thus
the relative energies of the spin polarized and unpolarized atoms are
evaluated on an equal footing with the other energy differences.

\begin{table*}
\caption{Basic structural and electronic properties of selected
elemental crystals.  The elements marked by $\dag$ (dagger) are not in
the experimentally determined crystal lattice but a simpler one; 
for these elements the numbers in the ``Expt.'' column are from the
all-electron calculations of Moruzzi {\em et
al.}~\cite{moruzzi.1978.cep}.  The elements marked by $*$ (asterisk)
are considered in the magnetic (spin polarized) ground state.}
\label{tab:elements}
       \begin{tabular}{c|c|ccc|ccc|ccc}
        Element & Crystal & & $\alpha_0$ (\AA) & & & B (GPa) & & & E$_{\text{coh}}$ (eV) & \\ 
                &         & LDA & GGA & Expt.       & LDA & GGA &
        Expt.    & LDA & GGA & Expt. \\
\colrule
        Li        & BCC     & 3.39 & 3.46 & 3.49  &   14.4 & 13.3 & 11.6  &    2.12 & 1.99 & 1.63  \\
        Be$^\dag$ & FCC     & 3.16 & 3.19 & 3.15  &   144 & 134 &  134    &    4.72 & 4.42 & 3.97   \\
        C         & DIA     & 3.53 & 3.53 & 3.57  &   464 & 451 & 443     &  8.89 & 8.13 &  7.37 \\
        Al        & FCC     & 4.12& 4.12  & 4.05  &   71.0 & 70.1 & 72.2  &   3.41 & 3.05 & 3.39  \\ 
        Si        & DIA     & 5.47 & 5.48 & 5.43  &   88   & 84.9 & 98.8  &   4.56 & 4.08 & 4.68 \\
        K         & BCC     & 5.17 & 5.32 & 5.23  &   4.7  &  3.9 & 3.2   &   1.05 & 0.96 & 0.93 \\
        Ca        & FCC     & 5.39 & 5.48 & 5.58  &   22.4 & 19.1 & 15.2  & 2.80 & 2.73 & 1.56  \\
        V         & BCC     & 3.01 & 3.06 & 3.03  &   198  &  169 & 162   &  7.32 & 5.67 & 5.31 \\
        Fe$^*$    & BCC     & 2.78 & 2.90 & 2.87  &   201  & 144 & 168    & 6.73 & 5.39 & 4.28  \\
        Ni$^*$    & FCC     & 3.48 & 3.59 & 3.52  &   248  & 179 & 186    & 5.70 & 4.43 & 4.44  \\
        Cu        & FCC     & 3.56 & 3.68 & 3.61  &   192  & 138 & 137    & 4.33 & 3.10 & 3.49  \\
        Ga$^\dag$ & FCC     & 3.92 & 3.92 & 4.14  &   75.6 & 73.7 & 44.0  &  3.22 & 2.89 & 3.22 \\
        Mo        & BCC     & 3.17 & 3.21 & 3.15  &   279  & 246 & 273    & 8.19 & 6.18 & 6.82 \\
    \end{tabular}
\end{table*}


\begin{figure}
\psfile{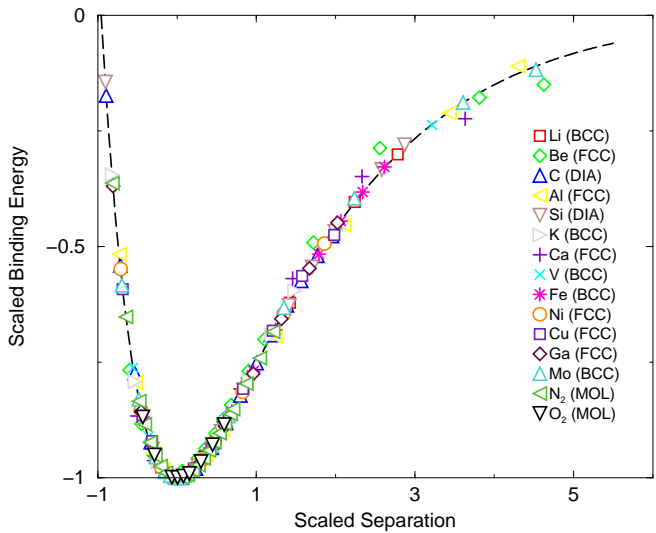}{1.00}
\caption{The bulk energies for various elemental solids and the N$_2$
and O$_2$ molecules, scaled using
the UBER\protect{\cite{rose.1984.ufe}}.  The
energies are calculated with the {\em a posteriori} gradient
correction to the LDA results.}
\label{fig:scaled-elemental}
\end{figure}

In Table~\ref{tab:elements}, we summarize the results of the
calculations. The experimental values are a compilation of results
from Kittel\cite{kittel.1996.iss}; for certain elements we considered
a simpler lattice than the experimental ground state, and in these
cases we compare our results to the all-electron calculations of
Moruzzi~\textit{et al}\cite{moruzzi.1978.cep}.  Note that, as
expected, the LDA results for the lattice constant are lower, and for
the bulk modulus are higher than the experimental values.  The
gradient correction tends to improve the situation.  Overall, the
agreement with experiment is quite good, except for the cohesive
energies, which is a well known deficiency of the approach we are
using.  Looking at the gradient corrected results, other than Ca which
seems to have large offsets, the lattice constants are within 2.0\% of
the experimental values, bulk moduli within 22\%, and cohesive
energies are within 26\%. In Fig.~\ref{fig:scaled-elemental} we
present the scaled results of the gradient corrected calculations and
the corresponding universal curve.  The magnetic moments of the
ferromagnetic materials at their equilibrium lattice constant values
as calculated with the LDA (GC) are 2.06 (2.30) $\mu_{\text{B}}$/atom
for Fe and 0.57 (0.59) $\mu_{\text{B}}$/atom for Ni.  These values
compare well with the experimental values of 2.22
$\mu_{\text{B}}$/atom for Fe and 0.61 $\mu_{\text{B}}$/atom for Ni, as
well as with the summary of the results of various
theoretical methods that is presented in
Ref.~\onlinecite{moroni:1997:upa}.

\subsection{Study of small molecules}
 \label{molecules}

In Sec.~\ref{bound_cond}, we described how two types of 
boundary conditions --- OBC and PBC --- can be readily 
used in a HARES calculation. 
Many DFT methods designed to do calculations for 
solids use PBC and are constrained to use a large supercell
to study an isolated molecule or a cluster of atoms. 
Here, we present results for four molecules N$_2$, O$_2$, H$_2$O, 
and NH$_3$ obtained using HARES with the two types of boundary
conditions keeping all other computational parameters fixed.
We use a periodic box of dimensions 
24 $\times$ 18 $\times$ 18 a.u.$^3$ for the N$_2$ 
and O$_2$ molecules and
20 $\times$ 20 $\times$ 20 a.u.$^3$ box for the 
H$_2$O and NH$_3$ molecules, with
a grid spacing of 0.25 a.u.

In Table~\ref{tab:molecule-BC}, the various results for small
molecules are summarized.  In ammonia, we also calculated the
inversion barrier of the potential energy surface, by relaxing the 
positions of the hydrogen atoms in the plane for selected heights 
of the Nitrogen atom.
We find that the bond-lengths obtained with OBC tend to be
smaller than those obtained with PBC, 
though the difference is quite small, in most cases smaller
than the accuracy in the reported results.
The energies, on the other hand, have significantly 
larger differences.
We suggest that this is due to the electrostatic interaction 
of the field in a supercell calculation with PBC.
This interaction between the molecule and its periodic 
images changes the energy of the molecule. In the 
examples considered here a cubic cell geometry is used,
which results in zero dipole interaction (it can be shown
analytically that the interaction energy of a dipole with
a full shell of dipoles is zero).  This indicates 
that the discrepancy is due to higher-order 
multipole terms.
The computational time for the calculations with PBC and OBC 
is similar, so there is no particular advantage to either approach
from the point of computational cost.
It appears, however, that for truly isolated systems 
the OBC approach gives more realistic results
due to the absence of any spurious long range fields.

\begin{table}
\caption{Calculated structure and energetics of 
         N$_2$, O$_2$, H$_2$O and NH$_3$: effect of boundary
conditions.  In ammonia the energy barrier 
for the inversion isomerization is given.  The
``Expt.''\  column contains the experimental values, the OBC
column has our results using the open boundary conditions, 
the PBC column has our results using the periodic boundary
conditions.} 
\label{tab:molecule-BC}
\begin{tabular*}{\columnwidth}{@{\extracolsep{\fill}}l|c|c|c|c}
\toprule
   Molecule &   Property  & Expt. &  OBC    &  PBC \\
\colrule
   N$_2$   &   Bond-length (\AA)      &  1.10   &   1.10  &  1.10   \\
           &   Cohesive Energy(eV)    &  9.9    &   9.8   &  10.0   \\
           &   Vibration (cm$^{-1}$)  &  2359   &   2395  &  2390   \\
   \hline
   O$_2$   &   Bond-length (\AA)      &  1.21   &   1.22  &  1.22   \\
           &   Cohesive Energy(eV)    &  5.20   &   7.63  &  7.79   \\
           &   Vibration (cm$^{-1}$)  &  1580   &   1584  &  1571   \\
   \hline
   H$_2$O  &   Bond-length (\AA)      &  0.98   &  0.97   &   0.97  \\
           &   Bond-angle             &  $104.6^{\circ}$  &  $105.7^{\circ}$  &  $104.4^{\circ}$  \\
   \hline
   NH$_3$  &   H-H distance (\AA)     &  1.64   &  1.67   &  1.68   \\
           &   N-H bond-length (\AA)  &  1.00   &  1.02   &  1.02   \\
           &   Energy Barrier (eV)    &  0.23   &  0.13   &  0.10   \\ 
\botrule
\end{tabular*}
\end{table}

\subsection{Electronic Structure of Blue Bronze, 
         K$_3$Mo$_{10}$O$_{30}$
\label{sec:bb}}

        The material called blue bronze (BB), 
whose chemical composition is A$_{0.3}$MoO$_{3}$ with A an alkali 
metal, exhibits a variety of interesting physical properties
including a metal-to-semiconductor transition at $T_c = 180$~K,
quasi-one-dimensional electronic properties above $T_c$, 
and the existence of incommensurate and commensurate charge density 
wave (CDW) phases~\cite{schlenker89}. 
Recently, a family of molybdenum bronzes has been extensively studied in
experiments using angle-resolved photoemission spectroscopy (ARPES) 
to explore a possible realization of non-Fermi-liquid behavior due to its
low-dimensional electronic properties~\cite{claessen95,gweon96,denlinger99}. 
To our knowledge, the only published electronic band calculation of 
BB is based on a tight-binding (TB) method using some model 
structures~\cite{whangbo86}. 
The dispersion of the TB bands around the Fermi level is 
\textit{qualitatively} different from the ARPES 
experimental results~\cite{gweon96}. 
It is of great importance to have an accurate and reliable 
\textit{ab initio} calculation of the electronic 
structure of BB in order to interpret the ARPES measurement
in terms of possibly interesting physics. With a large number of 
atoms in the unit cell, including 10 Mo atoms and 30 O atoms
which are typically difficult to handle with PW approaches,
BB provides a challenging system for performing state-of-the-art 
\textit{ab initio} calculations of the electronic structure; 
this requires a highly efficient computational tool such as HARES.

\begin{figure}
\psfile{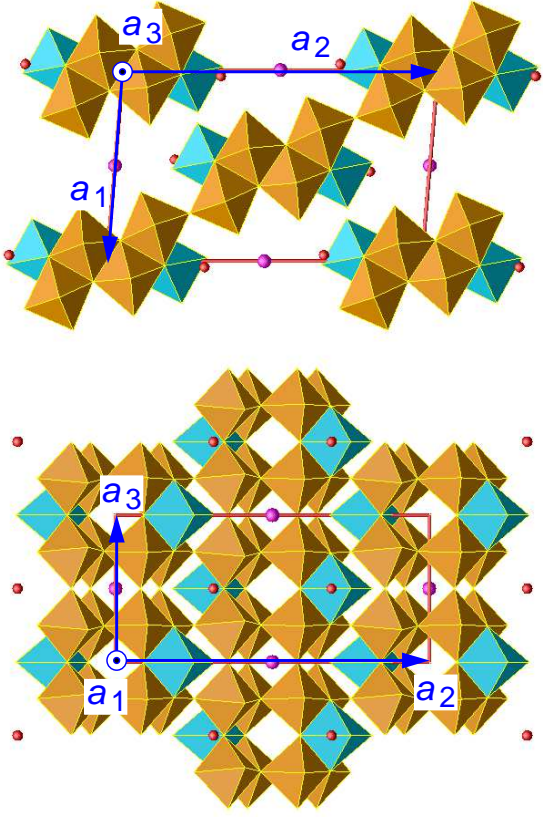}{0.9}
\caption{The crystal structure of K$_3$Mo$_{10}$O$_{30}$ depicted with 
         MoO$_6$ octahedra and K ions. The box indicates the simple 
         monoclinic 
         unit cell, red balls represent K ions, blue octahedra represent 
         the electronically
         inactive MoO$_6$ units, and yellow octahedra 
         represent the active MoO$_6$ units.
         Upper and lower panels correspond to top and front view,
         respectively.}
\label{fig:BB-str}
\end{figure}

The structure and the lattice parameters of BB is well documented 
in Ref.~\onlinecite{ghedira85}:
the Bravais lattice is centered monoclinic (CM), the space group is C2/m, 
and the lattice constants of the simple monoclinic cell 
are $a_2 = 16.2311$ \AA, $a_3 = 7.5502$ \AA, and $a_1 = 9.8614$ \AA\  
with the angle $\beta = 94.895^\circ$ 
between ${\mathbf a}_1$ and ${\mathbf a}_2$.
The basic building block of BB is the MoO$_6$ octahedron; ten octahedra
form a rigid unit by edge-sharing. Within the simple monoclinic (SM)
cell, 
two rigid units are arranged so that one of them is located at 
the apex and the other at the center of the cell. As a result,
neighboring rigid units share a corner oxygen to form a slab 
spanned by ${\mathbf a}_2- {\mathbf a}_1$ and ${\mathbf a}_3$ and
four infinitely-connected MoO chains (per primitive unit cell)
parallel to ${\mathbf a}_3$.
The crystal structure is illustrated in Fig.~\ref{fig:BB-str} where 
the simple monoclinic cell is indicated by a box; the two different classes
of octahedra are indicated by different colors: yellow for those that
participate actively to conduction along the high-conduction direction
(${\mathbf a}_3$) and blue for those that are apparently inactive.

We have performed electronic structure calculations for BB with HARES. 
For these calculations we use the Ceperley-Alder XC
functional as parametrized by Perdew and Zunger~\cite{CA,PZ-CA}.
The ions are represented by norm-conserving pseudopotentials generated
by the Troullier-Martins scheme~\cite{troullier91} in the fully separable form
of Kleinman and Bylander~\cite{kleinman,gonze}. 
The Brillouin zone (BZ) integrations are performed using a
$4 \times 4 \times 4$ Monkhorst-Pack
\textbf{k}-point mesh\cite{monkhorst.1976.spb} in the BZ of the CM cell.
We used a grid spacing of $h\simeq 0.33$~a.u.\ which gives around 200,000
real-space grid points.

\begin{figure}
\psfile{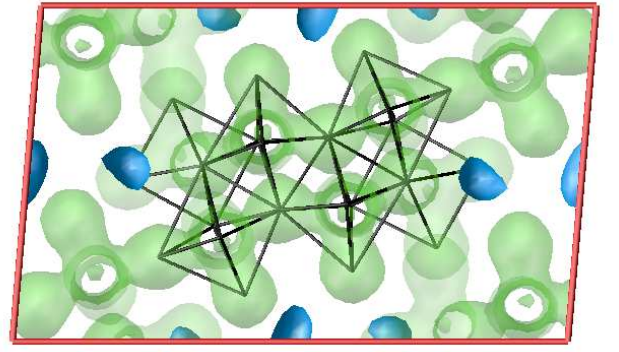}{1.00}
\caption{The isosurfaces of the valence electron density : the blue and
         the green surfaces correspond to a very low and a high
         density, respectively.}
\label{fig:BB-rho}
\end{figure}

In Fig.~\ref{fig:BB-rho}, we show isosurfaces of the valence electron 
density obtained from the fully self-consistent calculation. 
A few interesting features can be observed from the green
isosurfaces 
which correspond to high electron density:
\renewcommand{\labelenumi}{(\roman{enumi})}
\begin{enumerate}
\item the high density region has a cylindrical shape with its axes
      overlapping with MoO infinite chains; 
\item the electron density is ``marginally'' connected along 
      the [$\overline{1}10$] crystallographic direction,
      which is within the slab but perpendicular to the chain 
        direction; 
\item the electron density is ``barely'' connected along the 
        slab normal.
\end{enumerate}
From these observations, we expect that electronic conduction will be
highly anisotropic and the chain direction is the most favored.  On
the other hand, the blue isosurfaces, corresponding to low electron
density, indicate that the valence electrons are largely depleted
around the potassium ion sites, suggesting that K atoms play the role
of donors.

\begin{figure}
\psfile{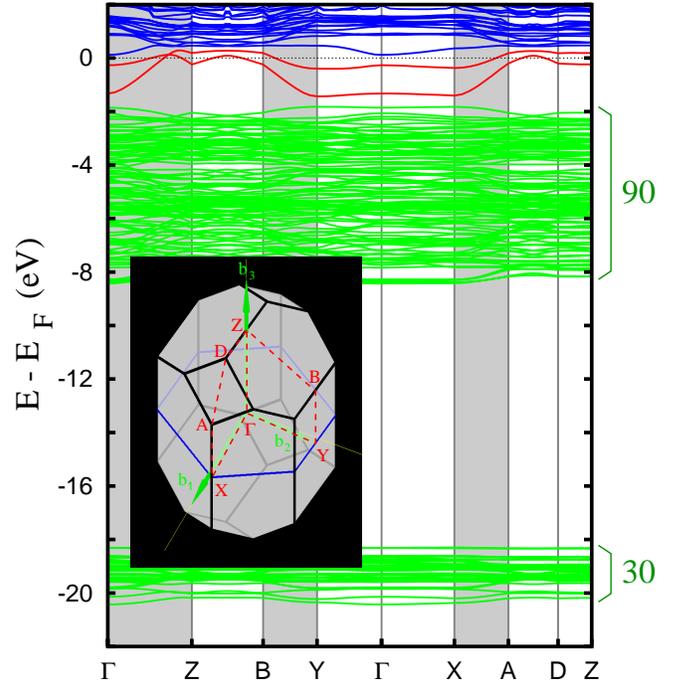}{1.00}
\caption{The \textit{ab initio} band structure of blue bronze: 
         Two bands cross the Fermi level and disperse
         significantly along the chain direction
         (indicated by gray panel).
         Notice the negligible dispersion in the other directions.
Inset is the Brillouin zone of the centered monoclinic cell and the
reciprocal lattice vectors ($b_1$, $b_2$, and $b_3$) of the simple
monoclinic cell are indicated.
}
\label{fig:BB-band}
\end{figure}

The full spectrum of the energy bands is shown in Fig.~\ref{fig:BB-band}.
The lower valence band manifold (30 bands) has O($2s$) orbital character, 
the upper valence band manifold (90 bands) has O($2p$) character 
mixed with Mo($4d$) orbitals near the top of the energy range. 
There are two bands crossing the Fermi level shown in red, well
separated from both the valence band and the conduction band manifolds.
These bands disperse primarily along the chain direction 
which is indicated by gray panels in Fig.~\ref{fig:BB-band}.

\begin{figure}
\psfile{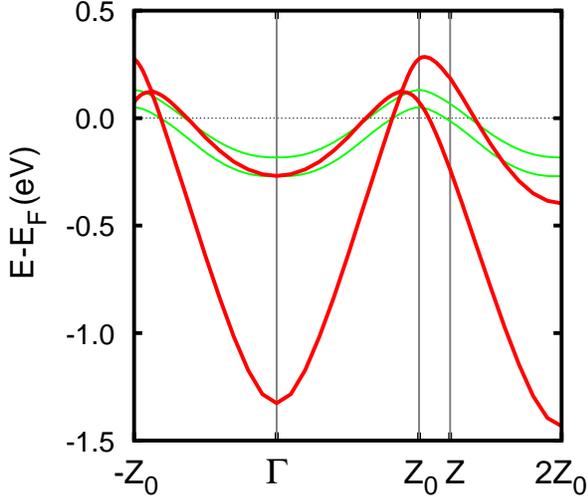}{1.00}
\caption{The {\em ab initio} bands (red) crossing the Fermi level are
compared with the tight-banding bands~\protect{\cite{whangbo86}} 
(green) in the neighborhood of
the Fermi level.  $z_0$ is a \textbf{k}-point on the simple monoclinic BZ
boundary and equal to $b_3/2$.}
\label{fig:BB-band-LDAnTB}
\end{figure}

The two partially filled bands of our \textit{ab initio} calculation
are qualitatively different from the TB bands~\cite{whangbo86}
 as illustrated
in Fig.~\ref{fig:BB-band-LDAnTB}.  For instance, the two LDA bands cross
each other near the BZ boundary whereas the TB bands do not.
Within the SM BZ (from -$z_0$ to $z_0$), the LDA bands have occupied
bandwidths of 1.3 and 0.3 eV while both of the TB bands disperse by
0.3 eV.  The occupied bandwidth of 1.3 eV for the low-lying band is
in good agreement with the APRES data\cite{gweon96}.
These qualitative differences in
the \textit{ab initio} and TB bands originate from the correct and
unbiased description of the interactions and the use of a realistic 
atomic structure in our calculations. Our analysis of the wavefunction
character at the $\Gamma$-point shows anisotropic hybridization between
the Mo $d$-states and the O $p$-states.  It would be rather difficult
to describe this situation within the context of the TB approach.

The significant dispersion of the partially filled bands only along the
chain direction results in planar Fermi surfaces, which are 
nested along the chain direction. In turn, the nested Fermi surfaces 
induce a CDW\@.  The estimated CDW wave vector is $\sim 0.75$ ${\mathbf b}_3$ 
compared with the observed one in the 
range of $0.72$ -- $0.75$ ${\mathbf b}_3$.
A more detailed analysis of the physics of this material will be
presented elsewhere~\cite{bb2000}. 

In summary, our application of HARES to the electronic structure of BB
suggests that an accurate and reliable method with a
realistic atomic structure is needed in order
to investigate 
the behavior of such complex materials.
The \textit{ab initio} energy bands are in good agreement with the ARPES
measurement and the nature of the electronic states relevant to conduction
can thus be elucidated. 

\subsection{Zeolite: Na$_n$Al$_n$ Si$_{24-n}$O$_{48}$
\label{zeolite}}

The word ``zeolite'' (of Greek origin) means ``boiling stone'' and derives 
from the visible loss of water when natural zeolite minerals are heated.
Zeolites are materials with unique properties which make them useful in a 
variety of applications such as oil cracking, nuclear waste management,
catalysis and animal feed supplements. They form a well-defined class of
naturally occurring crystalline alumino-silicate minerals. They have 
elegant three-dimensional structures arising from a framework 
of [SiO$_4]^{4-}$ and [AlO$_4]^{5-}$ coordination tetrahedra 
linked at their corners. 
The frameworks are generally very open and contain cavities that enclose
cations and water molecules.  
The presence of cavities make zeolites porous and
gives rise to their low density and unique properties. 
Since the cations,
water or other molecules that can be contained in these cavities, 
interact weakly with
the cavity walls, these entities have high mobility in the
solid zeolite. As a result a number of interesting physical and
chemical properties arise: facile ion exchange, easy water loss upon
heating, molecular sieve behavior, etc.

\begin{figure}
\psfile{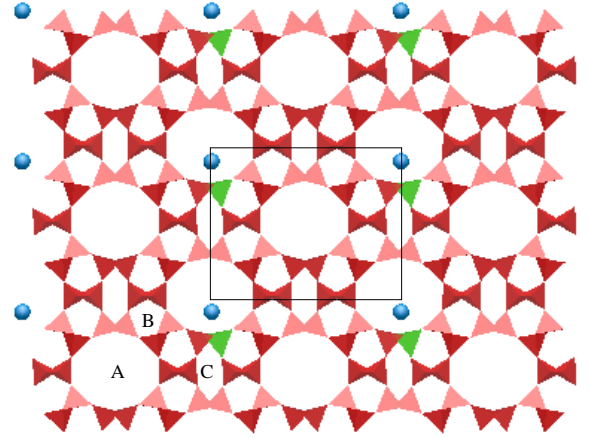}{1.0}
\caption{Framework in the structure of TON-zeolite, possessing three 
         types of cavities. O atoms occupy the positions at the 
         vertices of tetrahedra and Si (Al) atoms are at the center 
         of the red (green) tetrahedra.
         The Na atom (blue) remains inside the cavity in the vicinity 
         of the Al-occupied tetrahedron.  The rectangle in the center
denotes the projection of the orthorhombic unit cell perpendicular to
the $c$-axis.  In the bottom left corner we label the three different
types of pores.} 
\label{fig:zeo_str}
\end{figure}

In this Section, we study a zeolite which is referred to 
by the code name ``TON''~\cite{zeo_atlas} and has 
the general 
chemical formula Na$_n$Al$_n$ Si$_{24-n}$O$_{48}$; we will
consider the structures for $n=0$ and $n=1$. 
The framework of tetrahedra in its crystal structure is 
displayed in Fig.~\ref{fig:zeo_str}. TON has an orthorhombic 
crystal structure with three different types of pores or channels 
parallel to the $c$-axis. For $n=0$, its unit cell has 24 formula 
units of SiO$_2$ with a volume of about 1320~\AA$^3$. 
Its space group is Cmc2$_1$ and four of the 24 formula units 
are symmetry-independent. 
For $n=1$ (called Theta-1), the structure has been determined 
from X-ray powder experiments~\cite{acta_zeo}. Theta-1 is the first 
reported unidimensional medium-pore high-silica zeolite. 

Starting with the experimental geometry~\cite{acta_zeo}, 
we relaxed the atomic structure
of Si$_{24}$O$_{48}$ using HARES. 
All the bond lengths obtained from the calculation are
within 2\% of the experimental values. We next considered 
four independent Si sites where an Al atom can be 
substituted for a Si atom to obtain AlSi$_{23}$O$_{48}$, 
and relaxed its atomic structure. Interestingly enough, 
we find that all four possible structures have
very similar energies. Within the accuracy of our 
calculations the four sites cannot be differentiated.

\begin{figure}
\psfile{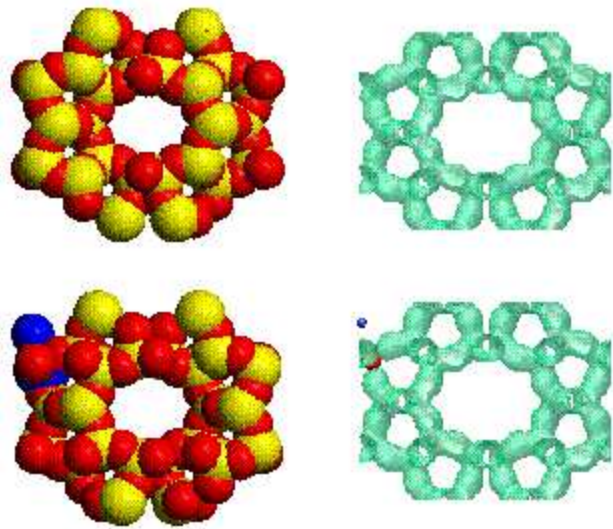}{1.0}
\caption{From left to right, top to bottom, we show the spacefill
structure model and the charge density isosurfaces of the
Si$_{24}$O$_{48}$
and the NaAlSi$_{23}$O$_{48}$ structures.  In the structure models 
the Oxygen atoms are red, the Silicon atoms are yellow.  In the
NaAlSi$_{23}$O$_{48}$ charge density plot blue corresponds to
Na, and red to Al.
        }
\label{fig:zeo_unit}
\end{figure}

Addition of a sodium atom to the AlSi$_{23}$O$_{48}$ structure 
introduces a variety of structural possibilities.  
We explored three possible structures, based on the three types 
of cavities in which the Na atom can be placed. 
In Fig.~\ref{fig:zeo_unit}, we show a unit cell of a structure with
Na added in the largest of the three cavities. 
Since AlSi$_{23}$O$_{48}$ is missing one electron due to the 
substitution of a Si atom by an Al atom, the added Na atom will 
naturally prefer to stay in the vicinity of the Al atom 
to which is can donate its valence electron. 
In the relaxed structure, the Na atom sits closest to three O atoms 
that are bonded to the Al atom and as a result the
Al-O bonds are elongated.

In Fig.~\ref{fig:zeo_unit}, we also show an isosurface of the 
electron density for NaAlSi$_{23}$O$_{48}$ and 
compare it with that of Si$_{24}$O$_{48}$. 
The two systems have very similar charge distribution
except in the region localized near the Na and Al atoms. 
The isosurfaces clearly have the same topology as the 
geometrical structure in Fig.~\ref{fig:zeo_unit}:
O atoms are at the center of the bulging regions 
of the isosurface and Si atoms at the joints or vertices. 
This indicates that the bonding is primarily ionic, 
with negatively charged O and positively charged Si atoms. 
Partial covalent character is also evident from the 
fully connected isosurface.

In NaAlSi$_{23}$O$_{48}$, the valence electron donated by 
Na compensates for the one missing electron in the four bonds 
formed by the  Al atom. 
The nature of bonding of Al with O atoms on the opposite 
side of the Na atom is clearly different from that 
with the three O atoms on the Na side.
The latter is very similar to the bonding character between 
Si and O in Si$_{24}$O$_{48}$. 
The charge on both Al and Na is positive, which has the effect 
of displacing the Al atom slightly away from the Na atom resulting
in longer Al-O bonds. This introduces small structural 
distortions and changes in charge distribution
in the neighboring SiO$_4$ tetrahedra. 
Since the addition of Na results in compensating 
electrostatic and covalent interactions, 
we expect that the energy barrier in the process of
attachment of Na (or in general a cation) to 
the walls of cavities in this zeolite should be very small. 
Further investigation of the chemical activity inside these 
pores and its effect in the electronic structure of the zeolite 
will be the subject of future studies~\cite{zeo2000}.

\section{Summary}
\label{sec:summary}

In this paper, we provided a comprehensive review of the theory 
underlying HARES, which is a method for 
{\it ab initio} electronic structure calculations implementated using
HPF on a shared memory 
parallel computer architecture. Several applications  
of the method to calculate the properties of simple and complex physical 
systems were presented to illustrate its capabilities. 
We obtained the bulk features of elemental solids such as equilibrium
lattice constant, bulk modulus and cohesive energy, for elements from
many different columns of the Periodic Table, and find good agreement
with experiment within the limitations of DFT/LDA calculations.
For the small molecules N$_2$, O$_2$, H$_2$O, and NH$_3$, we find that
the structural features do not depend on boundary conditions 
(open or periodic) used in the calculation, 
while the energy is sensitive to the the choice of 
boundary conditions. 
Application of the method to blue molybdenum bronze and
a zeolite demonstrate that it can be used effectively to study 
complex material systems.  In the case of      
Blue Bronze the results help to clarify important issues of the electronic 
structure pertaining to recent experiments. 
In our study of the TON zeolite Si$_{24}$O$_{48}$ and its variations 
containing Al and Na atoms, we demonstrated the ability of the method 
to capture the nature of bonding between a cation and the
walls of cavities in the zeolite; such interactions are related  
to the mobility of ions and molecules inside pores of the 
zeolite framework and should give rise to interesting physical 
and chemical behavior.

\section*{Acknowledgements}

The original development of the
adaptive-grid real-space  
method was supported by the 
Office of Naval Research 
through the Common High-performance Scientific Software Initiative (CHSSI)
and the High Performance Computation Modernization Office.
This set of codes is available upon request~\footnote{email:\tt
  kaxiras@cmt.harvard.edu}.   
The subsequent development of HARES was 
funded by Ryoka Systems Inc.
The authors wish to acknowledge useful discussions and collaborations with
Melvin Chen and Greg Smith, and useful comments
from Nick Choly, Ioannis Remediakis, Jose Soler and G.-H. Gweon.


\end{document}